\documentclass[aps]{revtex4}
\usepackage{amssymb}
\usepackage{color}
\usepackage[colorlinks=true,linkcolor=red]{hyperref}
\usepackage{hyperref}
\begin{document}
\title{Brane in 6D and localization of matter fields}
\author{Pavle  Midodashvili}
\email{midodashvili@hotmail.com} \affiliation{Department of
Physics,
  Tskhinvali State University,
  2 Besiki Str., Gori
   383500, GEORGIA}
\date{August 1, 2003}
\begin{abstract}
A new solution to Einstein's equations with a negative bulk
cosmological constant in infinite $(1+5)$-spacetime is found. It
is shown that the zero modes of all kinds of matter fields and
4-gravity are localized in $(1+3)$ subspace by the increasing and
limited from above warp factor.
\end{abstract}
\maketitle \indent{\ \ \ }  It is believed that the idea of extra
dimensions would be one of the most attractive ideas concerning
unification of gauge fields with general relativity. Lot of
attention has been devoted recently to alternatives of
Kaluza-Klein compactification . Large extra dimensions
\cite{RSh-Akama-Visser-RS,ADD} offer an opportunity for a new
solutions to old problems (smallness of cosmological constant, the
origin of the hierarchy problem, the nature of flavor, etc.). In
such theories our world can be associated with a $3$-brane,
embedded in a higher-dimensional spacetime with non-compact extra
dimensions and non-factorizable geometry. In this scenario, it is
assumed that all the matter fields are constrained to live on the
$3$-brane and the corrections to four dimensional Newton's gravity
low from bulk gravitons are small for macroscopic scales. But this
models still need some natural mechanism of localization of known
particles on the brane. The solutions to Einstein's equations in
extra dimensions and the questions of matter localization on the
brane have been investigated in various papers
\cite{Oda-GMSh-Giovan-CP-CK-BG-Pomarol-Greg-CW-Gogber,mg1-mg2,GRSh,G-Sh,R-DSh}.
In our opinion the localizing force must be universal for all
types of 4-dimensional matter fields. In our world the gravity is
known to be the unique interaction which has universal coupling
with all matter fields. If extended extra dimensions exist, it is
natural to assume that trapping of matter on the brane has a
gravitational nature. In the present paper our world is considered
as a single "fat" brane with the proper stress-energy tensor
embedded in $(1+5)$-spacetime. It must be noted that the
$(1+5)$-spacetime is of special importance to physics. That comes
from the observation that logarithmic gauge coupling unification
may be achieved in theories with two large spatial dimensions
\cite{A-HHSW}. The logarithmic behavior of the Green's functions
in effectively two dimensions has a chance of giving rise to
logarithmic variation of the parameters on the brane, thereby
reproducing the logarithmic running of coupling constants.
\\\indent{\ \ \ } In this article for the infinite (1+5)-spacetime
with a negative bulk cosmological constant we present the new
solution to Einstein's equations  with the increasing and limited
from above warp factor and show that the zero modes of all kinds
of matter fields and 4-gravity are localized on the $3$-brane.
\\\indent{\ \ \ } Let us begin with the details of our solution. In
6D the Einstein equations with a bulk cosmological constant
$\Lambda$ and stress-energy tensor $T_{AB}$  \begin{equation}
\label{6Dequations}R_{AB}  - \frac{1}{2}g_{AB} R = \frac{1}{{M^4
}}(\Lambda g_{AB}  + T_{AB} )\end{equation} can be derived from
the whole action of the gravitating system
\begin{equation}S = \int {d^6 x\sqrt { - g} \left[ {\frac{{M^4 }}{2}\left( {R +
2\Lambda } \right) + L} \right]}\end{equation}  $R_{AB}$ , $R$,
$M$ and $L$ are respectively the Ricci tensor, the scalar
curvature, the fundamental scale and the Lagrangian of  matter
fields (including brane). All of these physical quantities refer
to $(1+5)$- space with signature $(+ - ... -)$, capital Latin
indices run over $A,B,...=0,1,2,3,5,6$. Suppose that the equations
(\ref{6Dequations})  admit a solution that is consistent with
four-dimensional Poincar\'{e} invariance. Introducing for the
extra dimensions the polar coordinates $(r,\theta )$, where $0 \le
r <  + \infty$ , $0 \le \theta  < 2\pi$ , the six-dimensional
metric satisfying this ansatz we can choose in the form
\cite{GRSh} :
\begin{equation}\label{ansatzA} ds^2  = \phi ^2 (r)\eta _{\alpha \beta } (x^\nu  )dx^\alpha
dx^\beta   - dr^2  - g(r)d\theta ^2 ,\end{equation} where small
Greek indices $\alpha ,\beta ,... = 0,1,2,3$ numerate coordinates
and physical quantities in four-dimensional space, the functions
$\phi (r )$ and $g(r)$ depend only on  $r$  and are cylindrically
symmetric in the extra-space, the metric signature is $(+ - ...
-)$.The function $g(r)$ must be positive to fix the signature of
the metric (\ref{ansatzA}). \\\indent{\ \ \ } The source of the
brane is described by a stress-energy tensor $T_{AB}$ also
cylindrically symmetric in the extra-space. Its nonzero components
we choose in the form
\begin{equation}\label{branetensor}T_\nu ^\mu   = \delta _\nu ^\mu  F_0
(r),\ \ T_r^r  = T_\theta ^\theta   = F(r), \end{equation} where
we have introduced two source functions $F_0$  and $F$, which
depend only on the radial coordinate $r$. \\\indent{\ \ \ } By
using cylindrically symmetric metric ansatz (\ref{ansatzA})
 and stress-energy tensor (\ref{branetensor}),  the Einstein equations become
\begin{equation}\label{4dpartA} 3\frac{{\phi ''}}{\phi } + 3\frac{{\phi '^2 }}{{\phi ^2 }} +
\frac{3}{2}\frac{{\phi 'g'}}{{\phi g}} +
\frac{1}{2}\frac{{g''}}{g} - \frac{{g'^2 }}{{4g^2 }} =  -
\frac{1}{{M^4 }}(\Lambda  + F_0 ) + \frac{{\Lambda _{Phys}
}}{{M_P^2}},\end{equation}
\begin{equation}\label{55partA}6\frac{{\phi '^2 }}{{\phi ^2 }} + 2\frac{{\phi 'g'}}{{\phi g}} =  -
\frac{1}{{M^4 }}(\Lambda  + F) + 2\frac{{\Lambda _{Phys} }}{{M_P^2
}}\frac{1}{{\phi ^2 }} ,\end{equation}
\begin{equation}\label{66partA}4\frac{{\phi ''}}{\phi } + 6\frac{{\phi '^2 }}{{\phi ^2 }} =  -
\frac{1}{{M^4 }}(\Lambda  + F) + 2\frac{{\Lambda _{phys} }}{{M_P^2
}}\frac{1}{{\phi ^2 }},\end{equation} where the prime denotes
differentiation $d/dr$. The constant $\Lambda _{phys}$ represents
the physical four-dimensional cosmological constant, where
\begin{equation}\label{4deinstequationsA}R_{\alpha \beta }^{(4)}  -
\frac{1}{2}g_{\alpha \beta } R^{(4)}  = \frac{{\Lambda _{phys}
}}{{M_P^2 }}g_{\alpha \beta }.\end{equation} In this equation
$R_{\alpha \beta }^{(4)}$, $R^{(4)}$ and $M_P$ are
four-dimensional physical quantities: Ricci tensor, scalar
curvature and Planck scale. \\\indent{\ \ \ } In the case $\Lambda
_{phys}=0$  from the equations (\ref{4dpartA}), (\ref{55partA})
and (\ref{66partA}) we can find
\begin{equation}\label{sfconA} F' + 4\frac{{\phi '}}{\phi }\left( {F - F_0 } \right) = 0,
\end{equation}
\begin{equation}\label{cfconA}g = \delta \phi '^2 , \    \
\delta=const>0, \end{equation}\begin{equation}
\label{equationA}4\frac{{\phi ''}}{\phi } + 6\frac{{\phi '^2
}}{{\phi ^2 }} =  - \frac{1}{{M^4 }}\left( {F + \Lambda } \right)
.\end{equation} The (\ref{sfconA}) represents the connection
between source functions, it is simply a consequence of the
conservation of the stress-energy tensor and can be also
independently derived directly from $D_A T_B^A = 0$. \\\indent{\ \
\ } Suppose $F(r) = \varepsilon \phi ^2  + \rho \phi$, where
$\varepsilon$ and $\rho$ are some constants. Then taking the first
integral of the last equation \cite{mg1-mg2}, we get
\begin{equation}\label{equationAA}\phi '^2  = k\phi ^2 \left( {\frac{{5\varepsilon
}}{{7\Lambda }}\phi ^2  + \frac{{5\rho }}{{6\Lambda }}\phi  + 1}
\right) + \frac{C}{{\phi ^3 }}, \end{equation} where we have
introduced the parameter \begin{equation} k = - \frac{\Lambda
}{{10M^4 }}, \end{equation} and $C$ is the integration constant.
Setting $C=0$ and introducing boundary conditions at the origin
and at the infinity of transverse space
\begin{equation}\left. \phi  \right|_{r = 0}  \simeq 1,\ \ \left. \phi  \right|_{r =  +
\infty }  = const > 1,
\end{equation} in the case
\begin{equation}\Lambda  < 0,\ \ \varepsilon  < 0,\ \ \rho  = 12\sqrt
{\frac{{\varepsilon \Lambda }}{35}} ,
\end{equation} we can easily find the solution of the equation
(\ref{equationAA})
\begin{equation}\label{Solution}
\phi (r) = \frac{1}{{(\alpha  + e^{ - \sqrt k r} )}}
\end{equation} where the parameter $\alpha$ is small \begin{equation}\label{alpha}
    \alpha  = \sqrt {\frac{{5\varepsilon }}{7\Lambda }}  \ll 1.
\end{equation}
This solution gives the warp factor $\phi^2 (r)$ growing from
value $(1+\alpha)^{-1}$ at the origin and approaching a finite
value $\alpha^{-1}$ at infinity of transverse space.\\\indent{\ \
\ } With our metric ansatz (\ref{ansatzA}), the general expression
for the four-dimensional reduced Plank scale $M_{P}$, expressed in
terms of $M$, is
\begin{equation}\label{4pmass}M_P ^2  = M^4 \int_0^{ + \infty } {\frac{{\sqrt { - {}^6g}
}}{{\phi ^2 }}dr} \int_0^{2\pi } {d\theta }  = \frac{{2\pi
}}{3}\sqrt \delta  M^4 \alpha ^{ - 3} \left( {1 - \frac{{\alpha ^3
}}{{\left( {1 + \alpha ^3 } \right)}}} \right)
\end{equation}
In the case (\ref{alpha}) when  $\alpha$  is small, we have
\begin{equation}\label{4pmass1}
M_P ^2  =\frac{{2\pi }}{3}\sqrt \delta  M^4 \alpha ^{ -
3}.\end{equation} The inequality $M \ll M_P $ is possible by
adjusting $M^4 $ and the product of $\sqrt \delta$ by $\alpha ^{ -
3}$, and thus could lead to a solution of the gauge hierarchy
problem along the lines of \cite{ADD}.\\\indent{\ \ \ } A first
step in constructing a low energy effective theory is an analysis
of the spectrum of small perturbations around the solution
(\ref{Solution}). Consider the spectrum of small fluctuations of
metric, gauge and scalar fields and fermionic excitations near
background \cite{R-DSh} . An effective 4-dimensional low-energy
theory could arise if the following conditions are satisfied:
$(i)$ the spectrum contains normalizable zero (or small mass)
modes of graviton, gauge, scalar and fermion fields, with wave
functions of the type $\exp \left( {ip_\alpha x^\alpha  }
\right)\psi (r,\theta )$; $(ii)$ the effects of higher modes
should be experimentally unobservable at low energies.\\\indent{\
\ \ } We start from the spin-2 metric fluctuations $H_{\mu \nu }$
\begin{equation}ds^2  = \left\{ {\phi ^2 (r)\eta _{\alpha \beta } (x^\nu  ) +
H_{\alpha \beta } } \right\}dx^\alpha  dx^\beta   - dr^2  - \sqrt
\delta \phi '^2 d\theta ^2.\end{equation} From the 4-dimensional
point of view they are described by a tensor field $H_{\mu \nu }$
which is transverse and traceless:
\begin{equation}\label{graviton}\partial _\mu  H_\nu ^\mu   = 0,\ \ H_\mu ^\mu   =
0.
\end{equation}
This tensor is invariant under 4-dimensional general coordinate
transformations. For the fluctuations we can take the following
form\begin{equation} \label{fluctuations}H_{\alpha \beta }  =
h_{\alpha \beta } (x^\nu )\sum\limits_{ml} {\tau _m (r)\exp
(i\theta l)}.
\end{equation}
Separating the variables in the equations of motion for linearized
metric fluctuations (\ref{fluctuations})  it is easy to find the
equations for the 4-dimensional $h_{\alpha \beta }(x^\nu )$  and
the radial $\tau _m (r)$ modes
\begin{equation}\label{4dmodes}
\partial ^2 h_{\mu \nu } (x^\alpha  ) = m_0^2 h_{\mu \nu } (x^\alpha
),\end{equation}
\begin{equation}\label{radmodes}
\phi ^2(r) \tau ''_{m}(r) + \left( {2\phi(r) \phi '(r) +
\frac{{\phi ^2(r) \phi ''(r)}}{{\phi '(r)}}} \right)\tau '_{m}(r)
= m^2 \tau_{m}(r),
\end{equation} where
\begin{equation}\label{fluctmass}
m^2  = m_0^2  + \frac{{l^2 }}{{2\sqrt \delta  }}\frac{{\phi ^2(r)
}}{{\phi '^2(r) }}
\end{equation}
$m^2$ in (\ref{fluctmass}) contains the contributions from the the
orbital angular momentum $l$ in the transverse space. The
differential operator (\ref{radmodes}) is self-adjoint.It is easy
to see that the equation (\ref{radmodes}) has the zero-mass
$(m_0^2  = 0)$ and $s$-wave ($l = 0$) constant solution $\tau
_0=const$. Substitution of this zero mode into the
Einstein-Hilbert action leads to
\begin{equation}\label{gravityaction} S^{(0)}  = \tau _0^2 \int
{drd\theta \frac{{\sqrt { - {}^6g} }}{{\phi ^2 }}
 \int {d^4 x\sqrt { - \eta } \partial ^\beta  h^{\mu \nu } \partial _\beta
 h_{\mu \nu } }  + ...}=
\end{equation}
\begin{equation}
 =2\pi \sqrt \delta  \tau _0^2 \int\limits_{\left( {1 + \alpha }
  \right)^{ - 1} }^{\alpha ^{ - 1} } {\phi ^2 d\phi \int
  {d^4 x\sqrt { - \eta } \partial ^\beta  h^{\mu \nu }
  \partial _\beta  h_{\mu \nu } }  + ...}  = \end{equation}
  \begin{equation}\label{gravityaction1}= \frac{{2\pi \sqrt \delta  }}{3}\tau _0^2 \alpha ^{ - 3}
  \left( {1 - \frac{{\alpha ^3 }}{{\left( {1 + \alpha }
  \right)^3 }}} \right)\int {d^4 x\sqrt { - \eta }
  \partial ^\beta  h^{\mu \nu } \partial _\beta  h_{\mu \nu } }  + ... \ \ \ .\end{equation}The localization
  condition requires the integral over
$\phi$ in (\ref{gravityaction}) to be finite, as it is actually
(\ref{gravityaction1}). A wavefunction in flat space can be
defined as
\begin{equation}\label{zeromode}
\psi _0 (r) = \sqrt {\frac{{3k\alpha ^3 }}{{2\pi \sqrt \delta
\left( {1 - \frac{{\alpha ^3 }}{{\left( {1 + \alpha } \right)^3
}}} \right)}}} \frac{{e^{ - \frac{1}{2}\sqrt k r} }}{{\left(
{\alpha  + e^{ - \sqrt k r} } \right)^4 }}
 , \end{equation}
so the zero-mode tensor fluctuation is localized near the origin
$r=0$ and is normalizable. The contribution from the nonzero modes
will modify four- dimensional gravity.  It is easy to show
\cite{G-Sh} that this corrections are suppressed by $O\left[
{{\raise0.7ex\hbox{$1$} \!\mathord{\left/
 {\vphantom {1 {n^3 }}}\right.\kern-\nulldelimiterspace}
\!\lower0.7ex\hbox{${r^3 }$}}} \right]$  and their effect on  flat
$4D$-space is weak.\\\indent{\ \ \ } Now let us consider the
action of a massless real scalar field coupled to gravity:
\begin{equation}\label{scalarfield2}
S_\Phi   =  - \frac{1}{2}\int {d^6 \sqrt { - {}^6g} g^{AB}
\partial _A \Phi \partial _B \Phi },
\end{equation}
from which the equation of motion can be derived in the form:
\begin{equation}\label{scalarfield equation}
\frac{1}{{\sqrt { - {}^6g} }}\partial _M \left( {\sqrt { - {}^6g}
g^{MN} \partial _N \Phi } \right) = 0.
\end{equation}
In the background metric (\ref{ansatzA}) the equation of
motion(\ref{scalarfield equation}) becomes:
\begin{equation}\label{scalarfield equation1}
\phi ^2 \phi '\eta ^{\mu \nu } \partial _\mu  \partial _\nu  \Phi
- \partial _r \left( {\phi ^4 \phi '\partial _r \Phi } \right) -
\frac{{\phi ^4 }}{{\delta \phi '}}\partial _\theta ^2 \Phi  = 0.
\end{equation} Let us look for solutions of the form $\Phi \left( {x^M } \right) =
\upsilon \left( {x^\mu  } \right)\sum\limits_{l,m} {\rho _m \left(
r \right)e^{il\theta } }$, where the 4-dimensional scalar field
satisfies Klein-Gordon equation $\eta ^{\mu \nu } \partial _\mu
\partial _\nu \upsilon \left( {x^\alpha  } \right) = m_0^2
\upsilon \left( {x^\alpha  } \right)$, then for the radial modes
$\rho _m \left( r \right)$ we have
equation\begin{equation}\label{radialmodes1}\partial _r \left(
{\phi ^4 \phi '\partial _r \rho _m \left( r \right)} \right) =
\phi ^2 \phi '\left( {m_0^2  + \frac{{\phi ^2 }}{{\delta \phi '^2
}}l^2 } \right)\rho _m \left( r \right).
\end{equation} It is clear that this equation has the zero-mass ($m_0=0$) and s-wave ($l=0$) constant
solution $\rho _m \left( r \right)=\rho _0=const$. Let us
substitute the zero mode $\rho _0$ into the starting action
(\ref{scalarfield2}) and check if the constant solution is a
normalizable solution or not. With $\Phi _0 \left( {x^M } \right)
= \upsilon \left( {x^\mu  } \right)\rho _0$, the action
(\ref{scalarfield2}) can be cast to
\begin{equation}\label{scalarfieldzeromodeaction1}S_\Phi ^{(0)}  =  -
\frac{1}{2}\int {d^6 \sqrt { - {}^6g} g^{AB}
\partial _A \Phi _0 \partial _B \Phi } _0  =
\end{equation}
\begin{equation}
= - \frac{1}{2}\int {drd\theta \sqrt \delta  \phi ^2 \phi '\rho
_0^2 } \int {d^4 x\sqrt { - \eta }\eta ^{\nu \mu } \partial _\nu
\upsilon \left( {x^\alpha } \right)\partial } _\mu  \upsilon
\left( {x^\alpha  } \right) =\end{equation}
\begin{equation}\label{scalarfieldzeromodeaction2} =  - \pi \sqrt \delta  \rho
_0^2 \int\limits_{\left( {1 + \alpha }
  \right)^{ - 1} }^{\alpha ^{ - 1} }{\phi ^2 d\phi } \int {d^4 x\sqrt { - \eta }\eta ^{\nu
\mu }
 \partial _\nu  \upsilon \left( {x^\alpha  } \right)\partial } _\mu
 \upsilon \left( {x^\alpha  } \right) =\end{equation}
  \begin{equation}\label{scalarfieldzeromodeaction3}= - \frac{{\pi \sqrt \delta  }}{3}\rho _0^2 \alpha ^{ - 3} \left(
{1 - \frac{{\alpha ^3 }}{{1 + \alpha ^3 }}} \right)\int {d^4
x\sqrt { - \eta }\eta ^{\nu \mu } \partial _\nu  \upsilon \left(
{x^\alpha  } \right)\partial } _\mu  \upsilon \left( {x^\alpha  }
\right).\end{equation} As in the previous case the localization
condition requires the integral over $\phi$ in
(\ref{scalarfieldzeromodeaction2}) to be finite, as it is
actually. So the 4-dimensional scalar field is localized on the
brane.\\\indent{\ \ \ } Let us start with the action of $U(1)$
vector field:
\begin{equation}\label{vectorfield}S_F  =  - \frac{1}{4}\int {d^6 x\sqrt { - {}^6g} g^{AB} g^{MN}
F_{AM} F_{BN} },
\end{equation} where $F_{MN}  = \partial _M A_N  - \partial _N
A_M$ as usual. From this action the equation of motion is given by
\begin{equation}\label{vectorequationofmotion}
\frac{1}{{\sqrt { - {}^6g} }}\partial _M \left( {\sqrt { - {}^6g}
g^{MN} g^{RS} F_{NS} } \right) = 0.\end{equation} For the metric
(\ref{ansatzA}) this equation becomes
\begin{equation}\label{vectorequationofmotion1}
\phi ^2 \phi '\partial _\mu \left( {\eta ^{\mu \nu } g^{RT} F_{\nu
T} } \right) - \partial _r \left( {\phi ^4 \phi 'g^{RT} F_{rT} }
\right) - \frac{{\phi ^4 }}{{\delta \phi '}}g^{RT} \partial
_\theta  F_{\theta T}  = 0.\end{equation} By choosing the gauge
condition $A_{\theta}=0$ and decomposing the vector field as
\begin{equation}\label{decomposition1}A_\mu  \left( {x^M } \right) = a_\mu  \left( {x^\mu  }
\right)\sum\limits_{l,m} {\sigma _m \left( r \right)e^{il\theta }
},\end{equation}
 \begin{equation}\label{decomposition2}A_r \left( {x^M } \right) = a_r \left( {x^\mu  }
\right)\sum\limits_{l,m} {\sigma _m \left( r \right)e^{il\theta }
},\end{equation} it is straightforward to see that there is the
$s$-wave $(l=0)$ constant  solution $\sigma _m \left( r
\right)=\sigma _0=const$ and  $a_r=const$.  In deriving this
solution we have used $\partial _\mu  a^\mu   = \partial ^\mu
f_{\mu \nu }  = 0$ with the definition of $f_{\mu \nu }  =
\partial _\mu  a_\nu   - \partial _\nu  a_\mu$. As in the previous
cases, let us substitute this constant solution into the action
(\ref{vectorfield}). It turns out that the action is reduced to
\begin{equation}\label{vectorfieldconstant1}S_F^{(0)}  =  -
\frac{1}{4}\int {d^6 x\sqrt { - {}^6g} g^{AB} g^{MN} F_{AM}^{(0)}
F_{BN}^{(0)} }  =
\end{equation}
\begin{equation}\label{vectorfieldconstant2} =  - \frac{\pi }{2}\sqrt
\delta  \int\limits_{\left( {1 + \alpha } \right)^{ - 1} }^
{\alpha ^{ - 1} } {d\phi } \int {d^4 x\sqrt { - \eta } \eta ^
{\alpha \beta } \eta ^{\mu \nu } f_{\alpha \mu }^{(0)} f_{\beta
\nu }^{(0)} }  =
\end{equation}
\begin{equation}
= - \frac{{\pi \sqrt \delta }} {2}\alpha ^{ - 1} \left( {1
+\alpha} \right)^{-1}\int {d^4 x\sqrt { - \eta } \eta ^{\alpha
\beta } \eta ^{\mu \nu } f_ {\alpha \mu }^{(0)} f_{\beta \nu
}^{(0)} }.\end{equation} Thus, the vector field is localized on
the brane.\\\indent{\ \ \ } For spin $1/2$ fermion starting action
is the Dirac action given by
\begin{equation}\label{diracaction}S_\Psi   =
 \int {d^6 x\sqrt { - {}^6g} \overline \Psi  i\left(
{\Gamma ^M D_M  + m} \right)\Psi },\end{equation}  from which the
equation of motion is given by
\begin{equation}\label{diracequation}0 = \Gamma ^M D_M \Psi  =
 \left( {\Gamma ^\mu  D_\mu   + \Gamma ^r
D_r  + \Gamma ^\theta  D_\theta   + m} \right)\Psi .\end{equation}
An important feature of the action (\ref{diracaction}) is the
existence of mass term but we do not argue the origin of the term
at present. The relations between the curved-space gamma matrices
$(\left\{ {\Gamma ^A ,\Gamma ^B } \right\} = 2g^{AB})$ and the
minkowskian ones $(\left\{ {\gamma ^A ,\gamma ^B } \right\} =
2\eta ^{AB})$ read as follows:
\begin{equation}\label{curvedspacegammamatrices}\Gamma ^\mu
= \frac{1}{\phi }\gamma ^\mu  ,\ \ \Gamma ^r  = \gamma ^r , \ \
\Gamma ^\theta   = \frac{1}{{\sqrt \delta  \phi '}}\gamma ^\mu
.\end{equation} The spin connection is defined as
$$\omega _M^{\widetilde M\widetilde N}  = \frac{1}{2}h^{N\widetilde M}
\left( {\partial _M h_N^{\widetilde N}  - \partial _N
h_M^{\widetilde N} } \right) - \frac{1}{2}h^{N\widetilde N} \left(
{\partial _M h_N^{\widetilde M}  - \partial _N h_M^{\widetilde M}
} \right) -$$
$$-\frac{1}{2}h^{P\widetilde M} h^{Q\widetilde N}
\left( {\partial _P h_{Q\widetilde R}  - \partial _Q
h_{P\widetilde R} } \right)h_M^{\widetilde R},$$ where $\widetilde
M$, $\widetilde N$, ... denote the local Lorentz indices. The
non-vanishing components of the spin-connection for the background
metric (\ref{ansatzA}) are
$$\omega _\alpha ^{\overline r \overline \mu  }  = \phi '\delta
_\alpha ^{\overline \mu  } ,\ \ \ \ \omega _\theta ^{\overline r
\overline \theta  }  = \sqrt \delta  \phi ''.$$ Therefore, the
covariant derivatives have the form
$$D_\mu  \Psi  = \left( {\partial _\mu   + \frac{1}{2}\omega _\mu
^{\overline r \overline \alpha  } \gamma _{\overline r } \gamma
_{\overline \alpha  } } \right)\Psi ,\ \ \ D_r \Psi  = \partial _r
\Psi ,\ \ \ D_\theta  \Psi  = \left( {\partial _\theta   +
\frac{1}{2}\omega _\theta ^{\overline r \overline \theta  } \gamma
_{\overline r } \gamma _{\overline \theta  } } \right).$$  After
these conventions are set we can decompose the $6$-dimensional
spinor into the form $\Psi \left( {x^M } \right) = \psi \left(
{x^\mu  } \right)A\left( r \right)\sum {e^{il\theta } } $.  We
require that the four-dimensional part satisfies the massless
equation of motion $\gamma ^\mu  \partial _\mu  \psi \left(
{x^\beta  } \right) = 0$ and the condition $\Gamma ^r \psi \left(
{x^\mu  } \right) = \psi \left( {x^\mu  } \right)$. As a result we
obtain the following equation for the $s$-wave mode
\begin{equation}\label{swavediracmode}\left( {\partial _r  + 2\frac{{\phi '}}{\phi } +
\frac{1}{2}\frac{{\phi ''}}{{\phi '}}+m} \right)A\left( r \right)
= 0.\end{equation}  The solution to this equation reads:
\begin{equation}\label{diracsolution}A\left( r \right) = C\phi ^{ - 2} \phi '^{ - \frac{1}{2}} e^{ -
mr},\end{equation} with $C$ being an integration constant.
Substituting this solution into the Dirac action
(\ref{diracaction}) we have
\begin{equation}\label{spinorzeromodeaction1}S_\Psi ^{(0)}  = \int {d^6 x\sqrt { - {}^6g} \overline \Psi  _0
i\Gamma ^M D_M \Psi _0  = }
\end{equation}
\begin{equation}\label{spinorzeromodeaction2} = 2\pi \sqrt \delta  C^2 \int\limits_0^{ + \infty }
{\frac{{dre^{ - 2mr} }}{\phi }} \int {d^4 x\sqrt { - \eta }
\overline \psi  i\gamma ^\mu
 \partial _\mu  \psi  + ...} = \end{equation}
 \begin{equation}\label{spinorzeromodeaction4} = 2\pi \sqrt \delta  C^2 \left( {\frac{\alpha }{{2m}}
  + \frac{1}{{2m + \sqrt k }}} \right)\int {d^4 x\sqrt { - \eta } \overline \psi
   i\gamma ^\mu  \partial _\mu  \psi  + ...} \ \ \ . \end{equation}
In order to localize spin $1/2$ fermion in this framework, the
integral over $r$ in the equation (\ref{spinorzeromodeaction1})
should be finite. But this quantity is obviously convergent.
\\\indent{\ \ \ } To summarize, in this Letter we have presented new solution
to the Einstein's equations in the $(1+5)$ -spacetime. These
solution is found for the negative bulk cosmological constant
$\Lambda  < 0$ and it has increasing scale factor $\phi \left( r
\right)$ and a convergent volume integral, although the
transversal $2$-space is infinite. In addition for our solution we
have presented a complete analysis of localization of a bulk
fields at the origin in the extra space.

{\bf Acknowledgements:} Author would like to acknowledge the
hospitality extended during his visits at the Abdus Salam
International Centre for Theoretical Physics where the main part
of this work was done.

\end{document}